\documentclass[sigconf]{acmart}

\newcommand\black[1]{\color{black}#1}

\AtBeginDocument{%
  \providecommand\BibTeX{{%
    \normalfont B\kern-0.5em{\scshape i\kern-0.25em b}\kern-0.8em\TeX}}}

\setcopyright{acmcopyright}
\copyrightyear{2022}
\acmYear{2022}

 \acmConference[To appear in ACM CSCW '22]{Taipei '22}
\acmBooktitle{CSCW 2022}

\usepackage{comment}
\usepackage{subcaption}

\begin{document}

\title[Exploring the Impact of Rater Identity on Toxicity Annotation]{Is Your Toxicity My Toxicity? Exploring the Impact of Rater Identity on Toxicity Annotation}

\author{Nitesh Goyal}
\email{teshg@google.com}
\orcid{0000-0002-4666-1926}
\affiliation{%
  \institution{Google Research, Google}
  \streetaddress{111 8th Ave}
  \city{New York}
  \state{NY}
  \country{USA}
  \postcode{11201}
}
\author{Ian D.\ Kivlichan}
\email{kivlichan@google.com}
\orcid{0000-0003-2719-2500}
\affiliation{%
  \institution{Jigsaw, Google}
  \streetaddress{111 8th Ave}
  \city{New York}
  \state{NY}
  \country{USA}
  \postcode{11201}
}
\author{Rachel Rosen}
\email{rachelrosen@google.com}
\orcid{0000-0003-2927-1245}  
\affiliation{%
  \institution{Jigsaw, Google}
  \streetaddress{111 8th Ave}
  \city{New York}
  \state{NY}
  \country{USA}
  \postcode{11201}
}

\author{Lucy Vasserman}
\orcid{0000-0002-6938-0713} 
\email{lucyvasserman@google.com}
\affiliation{%
  \institution{Jigsaw, Google}
  \streetaddress{111 8th Ave}
  \city{New York}
  \state{NY}
  \country{USA}
  \postcode{11201}
}

\renewcommand{\shortauthors}{Goyal et al.}

\begin{abstract}

  Machine learning models are commonly used to detect toxicity in online conversations. These models are trained on datasets annotated by human raters. We explore how raters' self-described identities impact how they annotate toxicity in online comments. We first define the concept of {\black s}pecialized {\black r}ater {\black p}ools: rater pools formed based on raters' self-described identities, rather than at random. We formed three such rater pools for this study--specialized rater pools of raters from the U.S.\ who identify as African American, LGBTQ, and those who identify as neither. Each of these rater pools annotated the same set of comments, which contains many references to these identity groups. We found that rater identity is a statistically significant factor in how raters will annotate toxicity for identity-related annotations. Using preliminary content analysis, we examined the comments with the most disagreement between rater pools and found nuanced differences in the toxicity annotations. Next, we trained models on the annotations from each of the different rater pools, and compared the scores of these models on comments from several test sets. Finally, we discuss how using raters that self-identify with the subjects of comments can create more inclusive machine learning models, and provide more nuanced ratings than those by random raters.
  \\[2pt]
  \emph{Please be advised that this work contains examples of toxic and offensive content.}
 \end{abstract}



\maketitle

\section{Introduction}
Toxic language, defined as rude, disrespectful, or unreasonable language that is likely to make someone leave a discussion \cite{dixon2018annotation}, is a pervasive problem online. 
Detecting toxic language in online conversations using {\black m}achine {\black l}earning {\black (ML)} models is an active area of interest. However, models built to detect such online toxicity in conversations can be biased. Recent work has shown that some of the classifiers based on these models are more likely to label non-toxic language from minority communities as toxic compared to equivalent language from non-minority communities \cite{gomes2019drag,sap-etal-2019-risk}. For example, it has been shown that the Perspective API\footnote{\url{www.perspectiveapi.com}}, a publicly available API to detect toxicity in text, is more likely to predict high toxicity scores for comments written in African American English than for other comments \cite{sap-etal-2019-risk}. Likewise, \citet{gomes2019drag} has shown that non-toxic tweets from drag queens are more likely to be scored as toxic by the Perspective API than tweets by known white supremacists. 

As of today, ML models that define online content as toxic or not fall short of reaching the goal of being free from bias. As is evident from the examples above, when models fail, they do not fail equally for all identities. Some identity groups, already disenfranchised, are hurt even more than others because the prevalence of toxic comments directed towards all identity groups is not equal. Similarly, the presence of harder to detect microaggressions is likely unequal across groups. While this is not intentional, it does create inequity. ML models are based on huge datasets that have been trained {\black on data labeled} by human raters or annotators{\black , also known as} crowd workers. However, every human annotator cannot be fully aware of the intricacies of how different communities are being hurt using toxic language. Though certain phrases and words might seem innocuous to those outside an identity group, they may be known to be toxic to people who self-identify as that identity through their shared and lived experiences--or vice versa: some comments are perceived within identity groups to be non-toxic, but may be perceived by outsiders as toxic. What if we could ask those who self-identify with an identity group to rate content that was related to their community? This way, those who are likely to be targeted, and who would be best equipped to label the data, would be the ones to determine the ground truth for models that classify toxicity online. 
This paper continues to build upon research in this space of creating groups of annotators based on some differentiating factor(s) \cite{kocon2021offensive, akhtar2021whose, duzha2021hate, wich2021introducing}.
More specifically, we explore how raters from two relevant identity groups, African American and LGBTQ, label data that represents those identities, and whether their ratings vary from those provided by a randomly selected pool of raters who do not self-identify with these identity groups.

\section{Related Work}
Our work fits within a broader picture of trying to understand annotator identity or lived experience as a source of expertise, particularly within the context of abusive language and toxicity. We use the term ``annotator'' and ``rater'' interchangeably throughout this work to acknowledge that past literature has used both terms. 

We are not the first researchers who have been interested in understanding demographics of crowd workers and how that may impact crowdwork. For example, \citet{berg2015income} and \citet{posch2018characterizing} have asked questions about gender, age, and education level, among others. Further, \citet{posch2018characterizing} found that in the U.S., female crowdworkers are the majority, but in most other countries they are the minority. With respect to age, the authors found that most crowdworkers are between 18 and 34 years old; that remains consistent with previous findings \cite{berg2015income}.

We are also not the first to explore the impact of identity in building models for online conversation. Some papers have explored the identity of comment authors \cite{huang2020multilingual,subramanian2021evaluating}. In particular, \citet{halevy2021mitigating} explored author dialect as a proxy for author identity, fine-tuning a toxic comment classifier on African American English to reduce racial bias in the model. And by looking at identity on the community level, \citet{wich2020impact} explore{\black d} how conversations in different political communities have different norms. Our work expands {\black upon} this previous work by focusing on the identity of the annotators, instead of the authors.

Others have explored identity from the perspective of who the comment is targeting. \citet{kurrek2020towards} look{\black ed} at the usage of online slurs, in particular pejorative terms used against three distinct groups of people---gay men, {\black B}lack people, and transgender people. The authors have explicitly mentioned that due to the limitations of their research environment,  understanding relationships between annotator identity and annotations is a limitation of their work, something we explore in this paper. Others have also identified annotator bias as an area for future work \cite{huang2020multilingual,basile2021toward}. Our work expands \citet{kurrek2020towards} in {\black three} ways: {\black f}irst, by going beyond slurs and considering impact on multiple facets: toxicity, identity attacks, profanity, insults, and threats. Second, by annotating an entirely different dataset (Civil Comments instead of Reddit) by hundreds of carefully selected raters that belong to a specific identity group instead of 20 raters of mixed
identity groups. Third, \citet{kurrek2020towards} point out in Section 3.3 of their work that they could only perform limited analysis on the relationship between annotator demographics and annotations. They left this as an area for future work, which our work expands upon.

Other papers have considered the identity of annotators in groups, but along different lines than this work. For example, \citet{sap-etal-2019-risk} grouped annotators based on whether they have been primed to think about dialect and race. Authors found that annotators' different understandings of the same word in the same language could lead to racial biases in machine learning models when detecting hate speech. Our work expands
on these previous works by focusing on lived experiences of people instead of asking annotators to imagine those experiences.
Furthermore, \citet{waseem2016racist} grouped annotators based on domain expertise and found that amateur annotators (recruited without selection criteria) were more likely than experts (feminist and antiracist activists) to mislabel content as racist or sexist. 
This points to the importance of expertise, as expressed through the lived experiences and identity of the feminist and antiracist activists. Our work builds upon and differentiates in {\black a few} ways: we use and contribute an annotated version of the Civil Comments dataset instead of annotating Twitter{\black , and} we focus on lived identities of crowd workers and not the expertise of activists{\black , who} can be challenging to recruit{\black, making the work difficult to} scale. 
{\black Annotations by s}pecialized {\black rater}s, however, can be easily scaled up using various platforms that provide the ability to hire panels of raters based on several identity metrics. Perhaps most importantly, \citet{waseem2016racist} grouped feminist and antiracist activists together into one group, while we separate the identities into separate groups explicitly, showing nuanced differences between identities and annotations, i.e.\ that different identities are not the same and can not generically be grouped together.




Further, even when using experts who have deeper knowledge, annotating data appropriately is a challenge. In particular, \citet{davidson-etal-2019-racial} found biases even in the expert-annotated data. Authors found that that a classifier trained on expert-annotated data flagged Black-aligned tweets as sexist at nearly twice the rate of {\black W}hite-aligned tweets. Based on this, they cautioned that experts, like activists, may also hold biases similar to those of academics and crowdworkers, and emphasized the need for further work on exploring expertise and its role in the annotation process. In this work, we do not focus on expertise of trained experts because such experts are not always available and can be expensive to find and recruit. Instead we focus on crowd worker annotators and use their identit{\black ies} and lived experiences as a proxy for expertise.

Yet other works have grouped annotators into pools by forming clusters of annotators who rate similarly \cite{akhtar2020modeling,basile2020s,akhtar2021whose} or by using community-detection algorithms on a graph of annotators \cite{wich2020investigating}. Other have   incorporated a diversity of annotators in their rater pools but not explicitly grouped annotators according to identity and drawn conclusions based on identity groups specifically \cite{zeinert2021annotating}.




Identity effects on annotation have also been studied from the perspective of how gender can impact inter-annotator agreement, toxicity labels, and the resultant classifier performance \cite{binns2017trainer}. {\black \citet{wulczyn2017ex}} found that female annotators gave lower toxicity ratings on average than male annotators on the Wikipedia Detox dataset (note that there are different possible reasons for this), and analyzed which lexical features were weighted most heavily by classifiers trained on data labeled only by male or female annotators. More recently, broad comparisons of classifiers trained on data from demographically different annotator groups based on gender, first language, age, and education performed differently as measured by the classifier F1 score (harmonic mean of precision and recall) \cite{al-kuwatly-etal-2020-identifying}. They found that these features correlate with significant differences in classifier performance.

 Identity has also been recently explored by \citet{larimore-etal-2021-reconsidering}. Using a relatively small dataset, the authors found that White and non-White annotators rate tweets differently, especially in the context of certain topics (e.g.\ police brutality, antiracist politics, or empowering history), suggesting that the standards for evaluating racist language annotations should reflect the interpretations of those who are impacted or are at the receiving end. This lays the foundation for further research with larger data sets that can focus on particular identities and unpack their impact on ratings that power {\black m}achine {\black l}earning models that in turn are used to moderate content on the internet.

To summarize, understanding how demographics of annotators can impact annotations has been studied in seven different ways so far {\black in the literature:} 
\begin{enumerate}
    \item Creating rater pools \textit{a posteriori} by clustering raters based on their ratings and maximizing distances between these clusters instead of on the basis of their identities \cite{akhtar2020modeling, basile2020s, wich2020investigating}. 
    \item Creating identity-based pools on pre-existing datasets that looks for differences based on markers like age, gender, ESL, education e.g.\ on {\black the} Wikipedia {\black Detox Dataset} \cite{kocon2021offensive}. 
    \item Creating small, expert-based pools that perform annotations based on certain markers e.g.\ 3 annotators annotating immigrant/native status \cite{akhtar2021whose, duzha2021hate, wich2021introducing}. 
    \item As non-grouped random raters representing diversity to understand impact on annotations \cite{zeinert2021annotating, mohammad2021ethics}
    \item As identity-trained classifier-evaluation where classifiers trained on different pools of raters based on markers like gender, first language, age and education are evaluated as to how they perform in terms of precision and recall \cite{al-kuwatly-etal-2020-identifying}.
    \item As effects of comment authors' demographic differences and not the annotators' demographics as we study \cite{huang2020multilingual, yoder2021computational, liu2021authors, mosca2021understanding}, or creating new data sets based on political sub-communities of authors \cite{wich2020impact}.
    \item As focusing on model-robustness-evaluation to detect identity markers like race {\black and} gender correctly \cite{wang2021assessing}, multi-lingual hate speech classification \cite{aluru2020deep}{\black ,} and model-fairness-evaluation \cite{halevy2021mitigating, gencoglu2020cyberbullying}.
\end{enumerate}

Our work {\black was} motivated by three pieces of recent research and these works are perhaps the closest to ours. Work by \citet{basile2021toward} acts as the motivation for our work since it provides theoretical grounding of why data annotations matter for machine learning. Another close work {\black is} that by \citet{sap-etal-2019-risk}{\black ,}  where annotators are primed to think as if they might belong to specific ethnicit{\black ies}. While there is indeed value in asking annotators to consider putting themselves in the shoes of others, our work builds upon \citet{sap-etal-2019-risk} 's work by actually recruiting annotators that belong to {\black these} communities. Third, \citet{kurrek2020towards} presents work where the focus is on detecting slurs in Reddit data by a team of 20 annotators that are a mix of gender{\black s}, ethnicit{\black ies}, and sexual orientation{\black s}. The authors created a team of 20 annotators who represented inter-sectional identities across the three identity factors and they annotated slurs together as a team. 
Our work is a direct next step to this where we provide annotations on a different data set (Civil Comments dataset instead of Reddit), and we focus on {\black t}oxicity, {\black i}dentity {\black a}ttack{\black s}, {\black i}nsult{\black s}, {\black p}rofanity, and {\black t}hreat{\black s} instead of slurs, while showing relationships between annotators' ethnicit{\black ies} and sexual orientation{\black s} and their annotations. One of the goals of doing such annotations has also been to contribute this annotated dataset to the wider research community, {\black who} can use it to build models that can be used for classification and evaluation{\black ,} or perform deeper qualitative analysis.

This work has the following three primary contributions:
\begin{enumerate}
    \item We explore the impact on toxicity, insult, threat, identity attack, and profanity ratings of online conversations as perceived by two groups of annotators (referred to as {\black s}pecialized {\black r}ater {\black p}ools): African American and LGBTQ American raters
    \item \black{We are open-sourcing a large corpora of Civil Comments dataset and raters' annotations of conversations in this corpora across toxicity, insult, threat, identity attack, and profanity. We created and used this dataset to answer the above question and are sharing this dataset to encourage future research}
    \item We found that while {\black s}pecialized {\black r}ater {\black p}ools do create a statistically significant difference in annotations for online conversations as perceived by the two {\black s}pecialized {\black r}ater {\black p}ools, there are nuances to these differences, \black{highlighting pros and cons of using {\black s}pecialized {\black r}ater {\black p}ools.}
\end{enumerate}


\section{Methods}
\label{sec:methods}
\subsection{Research Questions}

Based on prior research, it is clear that individual raters will rate content differently. However, it is unclear whether raters will rate toxicity differently in comments written by or about their own identities than a randomly selected pool of raters who do not self-identify with these identities. Since we are focused on two particular communities, African Americans and LGBTQ Americans, we pose the following two Research Questions:

\begin{itemize}
    \item \textit{RQ1: Annotations by African American-identified raters on data related to the African American community will be measurably different from data annotated by raters who do not identify as African American }
    \item \textit{RQ2: Annotations by LGBTQ identified raters on data related to the LGBTQ community will be measurably different from data annotated by raters who do not identify as LGBTQ}
\end{itemize}

\subsection{Specialized and Control Rater Pools}
We define \textbf{specialized rater pools} as groups of raters that self-identify as specific identities, instead of a randomly selected group of raters chosen irrespective of their identities. Thus, for the purposes of this paper, we have two specialized rater pools: an African American specialized rater pool and an LGBTQ American specialized rater pool. {\black We chose to focus on these two identities because repeated bias with respect to these groups has been demonstrated in toxicity models previously \cite{sap-etal-2019-risk, gomes2019drag}. So,} to investigate our research questions, we ran three crowdrating tasks on a sample set of data with the following U.S.-based rater pools:

\begin{enumerate}
    \item A specialized rater pool with raters who identify as African American. 
    \item A specialized rater pool with raters who identify as  LGBTQ.
    \item A control rater pool with raters who identify as neither African American nor LGBTQ.
\end{enumerate}

One thing to note is the language used to describe the raters in rater pool (1); we are using the community description ``African American'' instead of ``Black'' to refer to this group. This is an intentional distinction; these group names, while often used interchangeably, are not identical--African American refers to a specific ethnicity within the Black community \cite{black-vs-aa}. Since previous work focused on African American English, we wanted to choose a group description that would be more likely to encompass speakers of that dialect, in order to build upon past work.

For the {\black c}ontrol rater pool (3), we alternatively considered having a rater pool consisting of raters selected at random irrespective of self-described identities. However, this would have led to interaction effects owing to identity groups present across both the control and specialized rater pools. Instead, the control rater pool consists only of raters who do not self-identify with either of the groups. This ensures that the rater pools remain disjoint. While we recognize that in a normal crowd-rating situation, raters of all identities are included at random, the goal of this paper is to understand if rater identity impacts toxicity annotation, and so we separate raters into distinct groups to test this question. 

\subsection{Designing Rater Pools: Crowd Contributor and Identity Considerations}

We worked with a third-party company to provide raters that perform annotation jobs. They recruited participants, on our behalf, to participate in this experiment. For the sake of this experiment, they managed constraints for recruiting raters, for which we specified age 18-35, English speaking, U.S. based raters.

Further, the third-party company provided a screener survey to select participants. We provided them with 50 screener questions and explanations for this purpose. These screener questions covered all aspects of toxicity and its subtypes, including identity attack, but we made sure to not include any ambiguous examples related to the experiment's identity groups in the screener questions {\black to avoid asserting correct answers for these cases}. Candidates taking the screener task saw the explanations when they got an item wrong, and were required to maintain an accuracy of 75\% or higher to be considered for further participation. The third party company chose participants with the highest accuracy for the task and placed them into rater pools based on their self-described identities. 

Raters with intersectional identities were chosen for a single rater pool rather than multiple rater pools so that the experimental groups remained disjoint. We did request that the third-party company keep the percentages of intersectional identities within each rater group within a tolerance of 1.5$\black \times$ the U.S. population averages (e.g.\ at the time the task was run, the LGBTQ percentage of the U.S.\ population was 4.5\% \cite{newport2018us}, and the African American population percentage was 13.4\% \cite{us-census-quick-facts}. So we requested the percentage of the LGBTQ rater pool that also identifies as African American should be no higher than 20.1\%, and the percentage of the African American rater pool that also identifies as LGBTQ should be no higher than 6.75\%, $1.5\times$ the population rates). 
This study has been approved by all internal review processes.

\subsection{Ethical Considerations when Working with Rater Identity}

There is a growing {\black recognition of the} importance {\black of considering the} ethical implications of how researchers perform work, especially with crowdworkers. In our context, raters perform the crowdwork of annotating potentially toxic text. We discuss {\black the ethical considerations we applied throughout} this work. 
\begin{enumerate}
    \item Minimize Data Leakage: As our third-party partner professionally manages raters, all of the three rater pools were constructed by them. They interfaced with the raters and had access to rater level information. We only received pseudonymized annotations that we analyzed. This was done to minimize leaking any identity-related information about the raters.
    \item Wage Fairness: We wanted to make sure that raters were paid fairly \cite{berg2015income}. While the raters are employed by the third-party vendor that we partnered with for the study, the vendor ensured us that raters were paid at least minimum wage for the jurisdiction. 
    \item Limit Toxic Content Exposure: Additionally, we restricted the toxicity of the data we asked for annotations, to limit participants' exposure to toxic content around their own identities. We did this by sampling 30\% of comments above a Perspective API toxicity threshold of 0.6, and the rest of the comments below that threshold, so that according to Perspective, they would be exposed to a maximum of 30\% toxic comments.
    \item Psychological Safety: Additionally, we consider{\black ed} raters' psychological safety. We gave raters the option to use a crowdrating platform-provided chatroom to communicate about the task so that they would have support, if needed. While this work involved limited engagement with toxic content, we still wanted to provide extra support.
    \item Upfront Transparency: We committed to transparency as recommended by \citet{kazimzade2020biased}. Since rater identity was being used to place raters into specialized pools, we informed raters about this. We also asked the vendor to communicate to raters that this is a study, so that we had rater consent to use the annotations for research purposes, since raters might otherwise expect the data was only for other common data annotation use cases, such as model building.
\end{enumerate}

\subsection{Dataset}
\label{sec:dataset}

The dataset used for this study is the Civil Comments dataset, which has been used previously for similar research about annotating toxic comments \cite{borkan2019nuancedmetrics}. The full Civil Comments dataset consists of approximately 2 million news comments from a now-defunct commenting platform. It has crowdsourced labels for toxicity and toxicity-subtypes, with 22\% of comments also labeled for identities. This data is public \footnote{\url{ https://www.kaggle.com/c/jigsaw-unintended-bias-in-toxicity-classification/data}} and the dataset itself is released under a CC0 license. It contains many references to identities, which is why we chose this dataset for this study.

From this dataset of 2 million news comments, we randomly sampled identity-neutral comments, as well as comments that mention each of the two identity groups in the study (African American and LGBTQ). The identity-neutral comments were sampled by excluding comments with identity labels (as provided by the dataset itself) for these identities. Next, we controlled for the rate of toxicity using Perspective API to mitigate the negative effects of toxicity exposure on raters, as we discussed in the Ethical Considerations section above. We recognize that Perspective API will make some errors here--this will affect our efforts to mitigate toxicity exposure, but will not otherwise affect the final results of the experiment, as raters will see each comment without knowing what the Perspective API score is or how the comment was sampled.

When sampling comments that mentioned each identity, we used a combination of identity labels as provided by the dataset as well as machine learning models that detect identity mentions in the text for the same identity labels used in the Civil Comments dataset. The comments that mentioned the identities were meant to serve as a proxy for both community-specific discussions and comments written about identities. Our sampling strategy means both are included, though it is challenging to distinguish between the two.

Overall, we created a dataset that contained a total of 25,500 comments from the Civil Comments dataset, with 8500 comments sampled to be identity agnostic, and 8500 comments sampled for each identity group (8500 for LGBTQ and 8500 for AA). The complete annotated data (including all individual annotations) {\black reflecting 382,500 annotations} is available on Kaggle in CSV and TSV formats as Google Specialized Rater Pools Dataset at \url{https://www.kaggle.com/datasets/google/jigsaw-specialized-rater-pools-dataset}  {\footnote{Nitesh Goyal, Ian Kivlichan, Rachel Rosen, \& Lucy Vasserman. (2022). [Data set]. Kaggle. https://doi.org/10.34740/KAGGLE/DSV/3533200}}. We hope that by releasing this data, we will enable the broader research community to build models using it to better understand the differences between the way annotators in the different pools annotate. Besides building new models,the community can also use this opportunity to dig deeper and perform content analysis on this datatset, which is beyond the goals of this paper. 

We next discuss the Likert scales and annotation template used for the task.

\subsection{Task Design}
All three rater pools were presented with the same full set of 25,500 comments, in a pre-sorted randomized order, for which we receive 5 ratings per annotator, a standard practice for similar work as shown by \citet{larimore-etal-2021-reconsidering}. Hence, the resulting dataset contained 15 ratings per comment: 5 ratings from annotators from each rater pool in the study. The task as seen by the raters uses the same template that has been used in previous works, for example by \citet{dixon2018annotation}. The task asks raters to rate comment toxicity, as well as other components of Toxicity 
on a Likert scale as first defined in \citet{dixon2018annotation} and defined inline below. 

\begin{enumerate}
    \item Toxicity is defined as ``a rude, disrespectful, or unreasonable comment that is likely to make people leave a discussion''. This is measured on a 4-point Likert scale with values between $-2$ and 1, where $-2$ = Very toxic, $-1$ = Toxic, 0 = Unsure, and 1 = Not toxic.

    \item Identity Attack is defined as ``negative or hateful comments targeting someone because of their identity''. This is measured on a 3-point Likert scale from $-1$ to 1.
    
    \item Insult is defined as ``insulting, inflammatory, or negative comment towards a person or a group of people''. This is measured on a 3-point Likert scale from $-1$ to 1.
    
    \item Profanity is defined as ``swear words, curse words, or other obscene or profane language''. This is measured on a 3-point Likert scale from $-1$ to 1.
    
    \item Threat describes ``an intention to inflict pain, injury, or violence against an individual or group''. This is measured on a 3-point Likert scale from $-1$ to 1.
\end{enumerate}


More details on the task itself are included in the Appendix: the full instructions are included in \autoref{fig:template}, and sample questions in \autoref{fig:template_examples}. 

\subsection{Measures}
\subsubsection{Descriptive Statistics}
\begin{enumerate}
    \item Toxicity Mean Difference: For each annotated comment, Toxicity Mean Difference is the difference in means of scores between each of the specialized rater pools and the control group, when the annotators rated the comment on {\black a} 4-point Likert scale as ``a rude, disrespectful, or unreasonable comment that is likely to make people leave a discussion''.
    \item Identity Comments with High Agreement: This is defined as the percentage of the comments that contain identity information and raters between control and specialized rater pools agree highly such that the Toxicity Mean Difference = 0.
    \item Identity Comments with Low Agreement: This is defined as the percentage of the comments that contain identity information and raters between control and specialized rater pools disagree highly such that the Toxicity Mean Difference $\ge1$.
 \end{enumerate}   

 \subsubsection{Regression Analysis}
\begin{enumerate}
    \item Toxicity Odds Ratio: This is the proportional odds for {\black s}pecialized {\black r}ater {\black p}ools to rate annotations as more likely to be toxic
    \item Identity Attack Odds Ratio: This is the proportional odds for {\black s}pecialized {\black r}ater {\black p}ools to rate annotations indicating higher likelihood of involving an identity attack
    \item  Insult Odds Ratio: This is the proportional odds for {\black s}pecialized {\black r}ater {\black p}ools to rate annotations as more likely to be insulting
    \item  Profanity Odds Ratio: This is the proportional odds for {\black s}pecialized {\black r}ater {\black p}ools to rate annotations as more likely to include profanity
    \item  Threat Odds Ratio: This is the proportional odds for {\black s}pecialized {\black r}ater {\black p}ools to rate annotations as more likely to be threatening
\end{enumerate}

\section{Results and Analysis}
\subsection{Descriptive Statistics}
\subsubsection{Rater disagreement}
\begin{figure*}[htp]
  \begin{subfigure}[a]{0.7\textwidth}
    \includegraphics[width=1\linewidth]{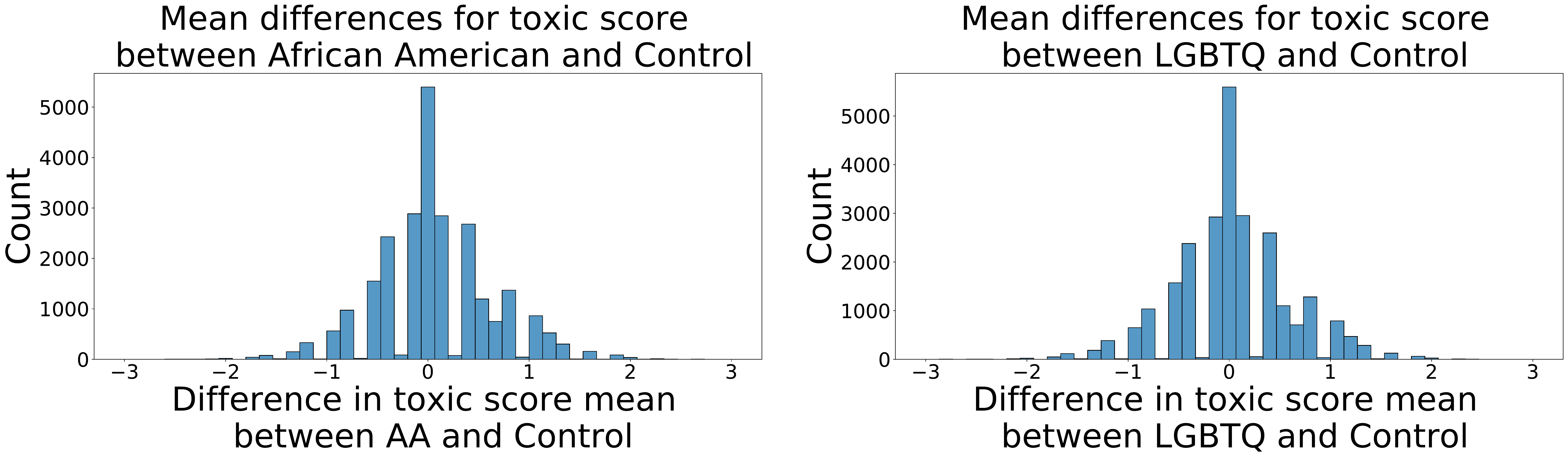}
    \caption{}
    \label{fig:pdf_toxic_score} 
  \end{subfigure}
  \begin{subfigure}[b]{0.7\textwidth}
     \includegraphics[width=1\linewidth]{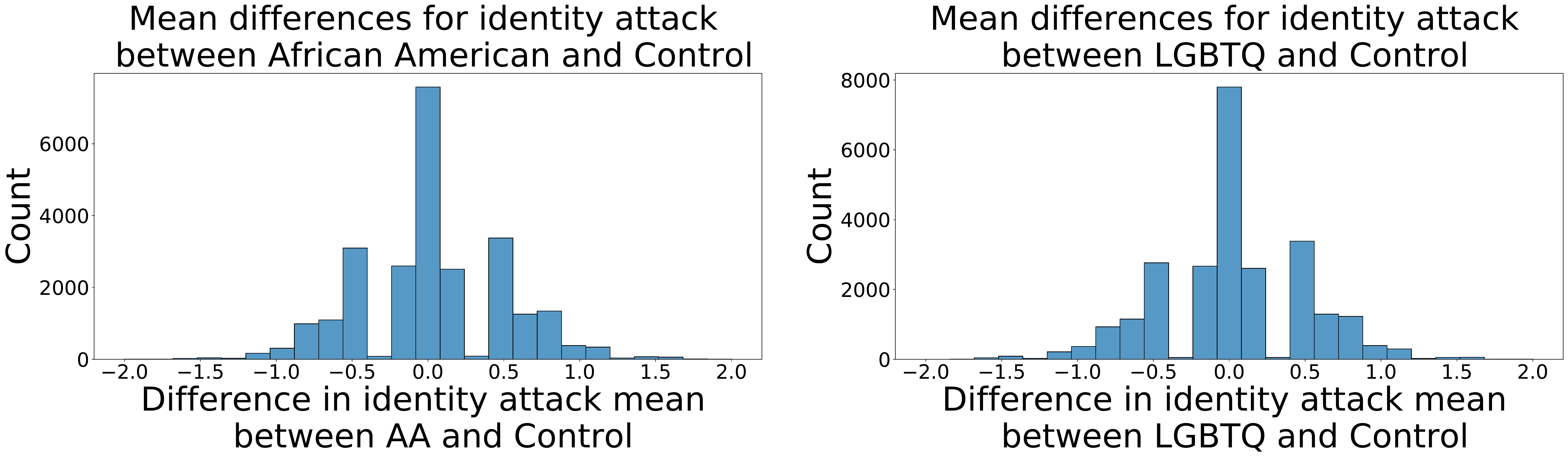}
     \caption{}
     \label{fig:pdf_identity_attack}
  \end{subfigure}
  \begin{subfigure}[c]{0.7\textwidth}
     \includegraphics[width=1\linewidth]{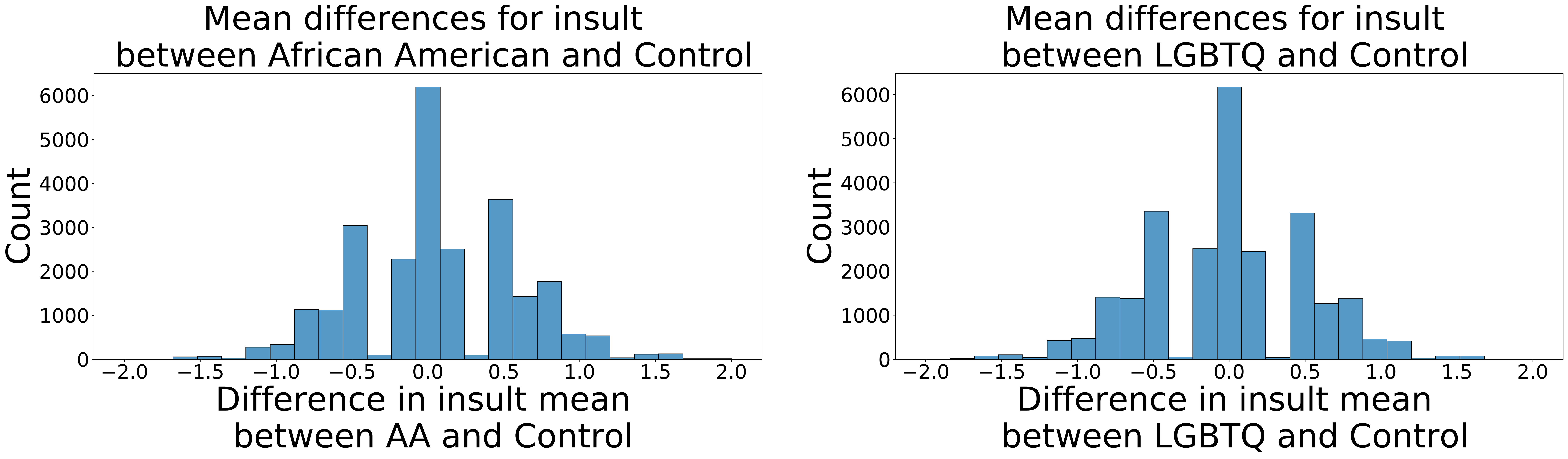}
     \caption{}
     \label{fig:pdf_insult}
  \end{subfigure}
  \begin{subfigure}[d]{0.7\textwidth}
     \includegraphics[width=1\linewidth]{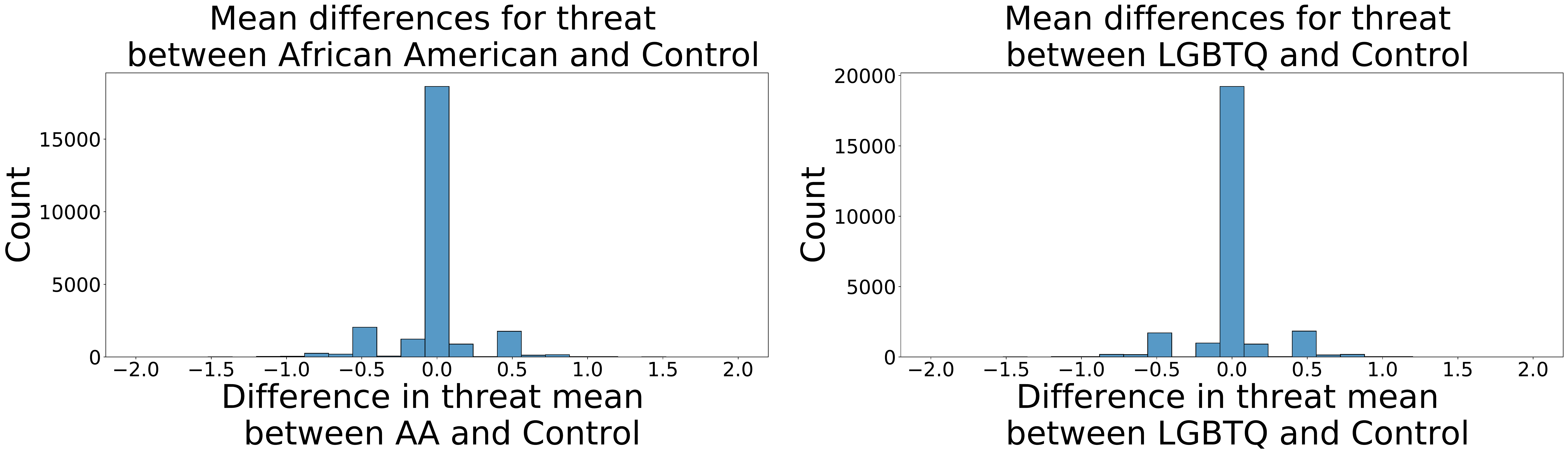}
     \caption{}
     \label{fig:pdf_threat}
  \end{subfigure}
  \begin{subfigure}[e]{0.7\textwidth}
     \includegraphics[width=1\linewidth]{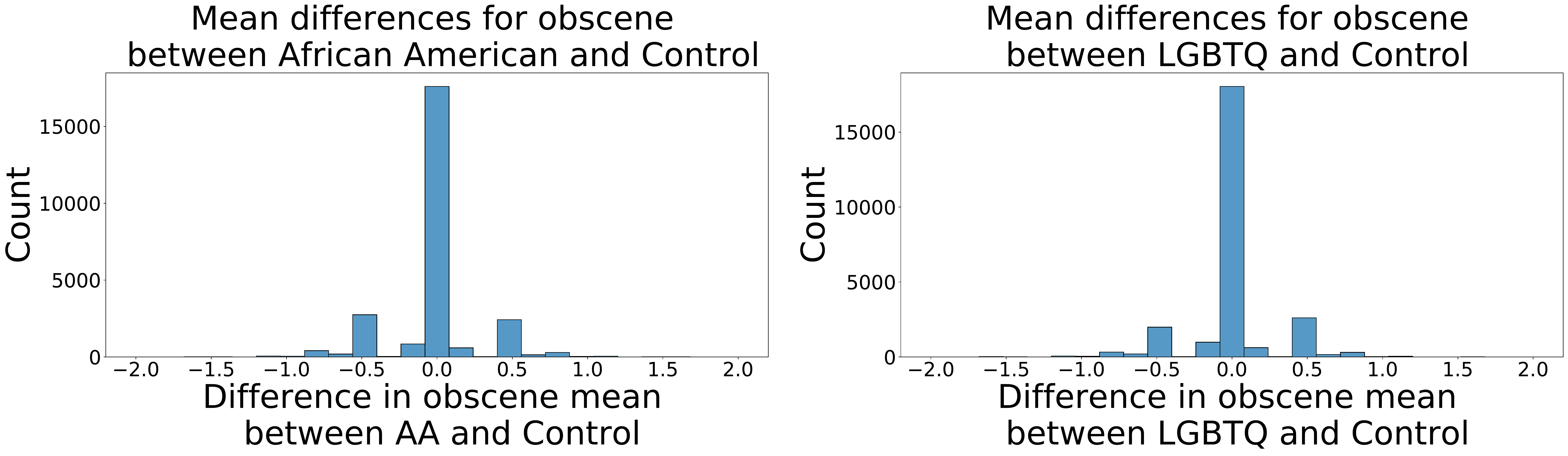}
     \caption{}
     \label{fig:pdf_obscene}
  \end{subfigure}
  \caption{{\black Histograms} of differences in scores, mean(specialized rater pool) - mean(control rater pool). The differences range from $-3$ (for the primary toxicity label; $-2$ for other labels), meaning that the specialized group rated the comment as much more likely to be toxic (or other label) than the control, to 3 (2 for other labels), meaning that the specialized group felt the comment was less likely to be toxic (or other label). {\black Histograms} for mean(African American) - mean(control) are on the left, and {\black histograms} for mean(LGBTQ) - mean(control) are on the right. (a) Represents the {\black histograms} for toxic score, (b) identity attack, (c) insult, (d) threat, and (e) profanity. Notably for all labels, the {\black histograms} are similar on the negative and positive sides, indicating that is not a trend towards more or less toxicity (other labels) among the specialized pools, the disagreements go in both directions. \label{fig:pdfs}}
\end{figure*}

{For each label, we consider the {\black histogram (counts)} 
where raters disagree. In \autoref{fig:pdfs} we show the probability distribution of the mean differences: mean(specialized rater pool) {\black $-$} mean(control rater pool), with negative differences meaning that the specialized pool rated the comment as more likely to be toxic (or {\black an}other label), and positive differences meaning the specialized pool rated the comment as less likely to be toxic. Notably for all labels, the distributions {\black of the histograms} are similar on the negative and positive sides, indicating that there is not a trend towards more or less toxicity ({\black or }other labels) among the specialized pools{\black ;} the disagreements go in both directions.

In \autoref{tab:percent_disagreement} we also explore the overall amount of disagreement between the rater pools, by showing the percentage of comments where the absolute value of the mean difference is greater than or equal to 1. We find that toxicity has the largest proportion of comments with disagreement (>12\% for both African American and LGBTQ rater pools), whereas the threat and profanity attributes have the least amount of disagreement, with <1\%.

We also consider how toxicity itself interacts with agreement between the rater pools by looking at the percentage of comments that are toxic (mean score < 0) among high and low disagreement comments. In \autoref{tab:toxic_disagreement_aa} we see these differences for the African American and control rater pools, and in \autoref{tab:toxic_disagreement_lgbtq} we see these differences for the LGBTQ and control rater pools. From this data we can see that according to all the rater pools, high disagreement comments have a higher percentage of toxicity than low disagreement comments, with the control rater pools overall finding more of these comments toxic than the specialized rater pools. 
}

\begin{table*}[ht]
  \caption{{\black The percentage of comments} where the specialized rater pool groups disagreed with the control rater pools. We see that toxicity has the largest proportion of comments with disagreement (>12\% for both African American and LGBTQ rater pools), whereas the threat and profanity attributes have the least amount of disagreement, with <1\%. \black{These are not to be read as stat. significant differences, but to highlight that some labels had more differences (Toxic Score, Identity Attack, and Insult) than others (Threat, and Profanity).}
  \label{tab:percent_disagreement}}
     \begin{tabular}{p{3cm} | p{3.5cm} | p{3.5cm} }
    \toprule
      Label  &  \% Comments with high AA-control disagreement &  \% Comments with high LGBTQ-control disagreement \\ 
   \midrule
     Toxic score  &  12.4\% &  12.5\% \\
     Identity attack &  5.7\% &  6.2\% \\
     Insult  &  8.5\% &  8.5\% \\
     Threat  &  0.5\% &  0.4\% \\
     Profanity  &  0.8\% &  0.8\% \\
   \bottomrule
  \end{tabular}
 \end{table*}

\begin{table*}[ht]
  \caption{The percent toxicity according to the African American and control rater pools. From the data, we can see that comments are more likely to be toxic according to both groups if there is high disagreement between the African American and control rater pools, with the control rater pool finding a higher percentage of comments to be toxic overall (7.2\%) than the African American rater pool (4.5\%). \black{These are not to be read as stat. significant differences, but to highlight that AA and control pools think of Toxicity differently}.}
  \label{tab:toxic_disagreement_aa}
     \begin{tabular}{p{5cm} | p{3cm} | p{3cm} }
    \toprule
     &  High AA-control disagreement 
      &  Low AA-control disagreement 
     \\
   \midrule
     \% of Comments that are Toxic 
    according to African American pool &   4.5\% &  2.3\% \\
     \% of Comments that are Toxic 
    according to {\black c}ontrol pool &   7.2\% &  2.3\% \\
   \bottomrule
  \end{tabular}
 \end{table*}
 
 \begin{table*}[ht]
  \caption{ Percentage of comments that are toxic according to the LGBTQ and control rater pools. From the data, we can see that comments are more likely to be toxic according to both groups if there is high disagreement between the LGBTQ and control rater pools, with the control rater pool finding a higher percentage of comments to be toxic overall (6.6\%) than the LGBTQ rater pool (5.3\%). \black{These are not to be read as stat. significant differences, but to highlight that LGBTQ and control pools think of Toxicity differently}.}

  \label{tab:toxic_disagreement_lgbtq}
     \begin{tabular}{p{5cm} | p{3cm} | p{3cm} }
    \toprule
 &  High LGBTQ-control disagreement 
      &  Low LGBTQ-control disagreement 
     \\
     
   \midrule
     \% of Comments that are Toxic 
    according to LGBTQ pool  &  5.3\% &  2.4\% \\
     \% of Comments that are Toxic 
    according to {\black c}ontrol pool  &  6.6\% &  2.4\% \\
   \bottomrule
  \end{tabular}
 \end{table*}
 
   \begin{table*}[ht]
  \caption{Percentage of comments that mention identities for different levels of agreement between African American/LGBTQ rater pools and the control rater pool. Comments with {\black no dis}agreement (mean difference = 0) have lower percentages of identity mentions than comments with {\black higher dis}agreement between groups (mean difference $\ge\!1$).}
  \label{tab:identity_disagreements}
  \begin{tabular}{p{3.6cm} p{2cm} p{2cm} p{2cm} p{2cm}}
    \toprule
    & High AA-control disagreement & No AA-control disagreement & High LGBTQ-control disagreement & No LGBTQ-control disagreement \\
    \midrule
    \% identity-mentioning comments & 71.0 & 49.5 & 71.0 & 49.4 \\
    \bottomrule
  \end{tabular}
 \end{table*}

 \subsubsection{Comments mentioning identity}
In Table \ref{tab:identity_disagreements} we examine the percentage of comments that mention identities for different levels of agreement between each of the African American and LGBTQ specialized rater pools and the control pool. We can see that comments with high disagreement between the specialized rater pools and the control pools tend to mention identities more often (71\% of the time) compared to comments with {\black no dis}agreement (50\% and 49\% of the time). This supports the hypothesis that specialized rater pools are critical for understanding comments that reference identity groups, since annotators in specialized rater pools are more likely to disagree with control group raters on these comments.

\subsection{Regression Analysis}
 Since the outcome variables for Regression Analysis are not binary (4{\black-}point Likert scale or 3{\black-}point Likert scale), we ran an ordinal logistic regression analysis to evaluate RQ1 and RQ2 \cite{agresti2003categorical, pregibon1981logistic}. We used the SPSS Advanced Statistics module to created dummy variables for recoding the dependent variables (Toxicity Score, Identity Attack, Insult, Profanity and Threat) and used {\black the} PLUM and GENLIN  commands to perform the analysis. This test has 4 assumptions and all of them were met: (a) dependent variable (Toxicity Score, Identity Attack, Insult, Profanity and Threat) should be ordinal; (b)  one or more independent variables that are continuous or categorical (Rater Pool category: control, LGBTQ, {\black African American}); (c) There should be no multicollinearity if there are two or more continuous independent variables (we only have one non-continuous variable: Rater Pool category); (d) There should be {\black p}roportional odds. Using separate Binomial Logistic Regressions on the dummy variables, we found that variables in the {\black e}quation had similar $\black \exp (B)$ values across the 3 dummy variable tests for all the measures, {\black where $\exp$ denotes the exponential and $B$ is the coefficient estimate}. Subsequently, a cumulative odds ordinal logistic regression with proportional odds was run {\black as a final model} to determine the effect of rater pool {\black as a independent variable}, with comment id and rater id as covariates, on identifying dependent variables {\black Toxicity Score, Identity Attack, Insult, Profanity and Threat}. The deviance goodness-of-fit test indicated that the model was a good fit to the observed data but most cells were sparse with zero frequencies.

\subsubsection{Toxicity Odds Ratio}
The final model statistically significantly predicted the dependent variable over and above the intercept-only model, 
$\chi^2(3) = 624.02$, $p < .001$. {\black The final model didn't code control as the intercept since the control involved random participants}. The odds of the {\black c}ontrol pool rating comments to be toxic was 0.957 (95\% CI, 0.939 to 0.976) times that of LGBTQ rater pool, a statistically significant effect, $\chi^2(1) = 19.88$, $p < .001$.
The odds of the {\black c}ontrol pool rating comments to be toxic was 0.986 (95\% CI, 0.968 to 1.004) times that of the {\black African American} rater pool, {\black not} a statistically significant effect, $\chi^2(1) = 2.27$, $p = .132$. This indicates that the control pool was slightly less likely to rate comments as toxic than {\black the LGBTQ pool, with no statistically significant difference between the {\black c}ontrol and the {\black African American} rater pool.}

\subsubsection{Identity Attack Odds Ratio}
The final model statistically significantly predicted the dependent variable over and above the intercept-only model, $\chi^2(3) = 371.51$, $p < .001$. The odds of the {\black c}ontrol pool considering comments to be identity attacks was 0.905 (95\% CI, 0.886 to 0.925) times that of the LGBTQ rater pool, a statistically significant effect, $\chi^2(1) = 80.15$, $p < .001$. The odds of the {\black c}ontrol pool considering comments to be identity attacks was 0.942 (95\% CI, 0.923 to 0.962) times that of the {\black African American} rater pool, a statistically significant effect, $\chi^2(1) = 31.333$, $p <.001$. Similar to toxicity, this indicates that the control pool was slightly less likely to rate comments as identity attacks than the specialized pools.

\subsubsection{Insult Odds Ratio}
The final model statistically significantly predicted the dependent variable over and above the intercept-only model, $\chi^2(3) = 715.37$, $p < .001$. The odds of the {\black c}ontrol pool considering comments to have to be insults was 0.930 (95\% CI, 0.912 to 0.948) times that of the LGBTQ rater pool, a statistically significant effect, $\chi^2(1) = 53.07$, $p< .001$. The odds of the {\black c}ontrol pool considering comments to be insults was 1.066 (95\% CI, 1.046 to 1.086) times that of the {\black African American} rater pool, a statistically significant effect, $\chi^2(1) = 45.113$, $p <.001.$ Here we see that the control pool was again less likely to rate comments as insults than the LGBTQ rater pool, but was more slightly likely to rate comments as insults than the {\black African American} pool.

\subsubsection{Profanity Odds Ratio}
The final model statistically significantly predicted the dependent variable over and above the intercept-only model, $\chi^2(3) = 116.539$, $p < 0.001$. The odds of the {\black c}ontrol pool considering comments to be profanity was 0.957 (95\% CI, 0.932 to 0.982) times that of the LGBTQ rater pool, a statistically significant effect, $\chi^2(1) = 10.74$, $p = .001$. The odds of the {\black c}ontrol pool considering comments to be profaned was 0.954 (95\% CI, 0.930 to 0.978) times that of the {\black African American} rater pool, a statistically significant effect, $\chi^2(1) = 13.92$, $p < .001$. Again, the control pool is less slightly likely to rate comments as profanity than both specialized pools.

\subsubsection{Threat Odds Ratio}
The final model statistically significantly predicted the dependent variable over and above the intercept-only model, $\chi^2(3) = 229.044$, $p < 0.001$. The odds of the {\black c}ontrol pool considering comments to be threats was 0.820 (95\% CI, 0.784 to 0.858) times that of the LGBTQ rater pool, a statistically significant effect, $\chi^2(1) = 75.325$, $p = .0001$. The odds of the {\black c}ontrol pool considering comments to be threats was 0.714 (95\% CI, 0.684 to 0.745) times that of the {\black African American} rater pool, a statistically significant effect, $\chi^2(1) = 13.92$, $p < .001$. Again the control pool is less likely to rate comments as threats than both specialized pools.

To summarize, these results mean:
\begin{enumerate}
\item RQ1:
\begin{itemize}
  \item Despite relatively small differences, ratings by the control pool were statistically significantly 
less likely to be tagged as identity attacks, profanity, and threats compared to ratings by the African American rater pool
  \item Despite relatively small differences, ratings by the control pool were statistically significantly more likely to be tagged as insults compared to ratings by the African American rater pool
  \item Ratings by the control pool were not statistically significantly less likely to be tagged as  toxic compared to ratings by the African American rater pool
\end{itemize}
\item RQ2:
\begin{itemize}
  \item Despite relatively small differences, ratings by the control pool were statistically significantly 
less likely to be tagged as toxic, having identity attacks, including insults, profanity, and threats compared to ratings by the LGBTQ rater pool
\end{itemize}
\item Disagreements:
\begin{itemize}
    \item {\black The c}ontrol pool disagrees more with the LGBTQ pool about identity attacks as compared to disagreeing about identity attacks with the African American pool
    \item {\black The c}ontrol pool rates more comments to be toxic than perceived by both the African American and LGBTQ pools, when there is significant disagreement
\end{itemize}
\end{enumerate}

\subsection{Preliminary Content Analysis}

\begin{table*}[ht]
  \caption{Sample comments with themes closely tied to identity where the African American and control rater pools had high disagreement (high differences in mean scores). We present here the fraction of raters who rated the comment as ``toxic'' or ``very toxic'', {\black $-1$ or $-2$} on the Likert scale respectively. These fractions are what we use for model training, which makes them comparable to Perspective's scores for these comments.}
  \label{tab:example_disagree_comments}
  \begin{tabular}{p{8cm} | p{1.4cm} | p{1.3cm} | p{1.3cm}}
    \toprule
     Comment & Perspective Toxicity score & Fraction of {\black c}ontrol Pool Raters who rated this toxic & Fraction of AA Pool Raters who rated this toxic  \\
    \midrule
    ``In other words, he's not brown enough for ya to be labeled a terrorist.'' & 0.56 & 1 & 0 \\
    ``I'm sure it was merely an oversight but...not mentioned in the story is that the killer was black and the victims were white. Jus' sayin'.'' & 0.38 & 0 & 1 \\
    ``This guy waved a flag of treason and then shot two cops. Let me guess, he's ``mentally ill,'' not a terrorist, because he's white.'' & 0.78 & 1 & 0 \\
    ``You apparently can say whatever you want about Mexicans, Hispanics \& Black people, but the Republican Party draws the line on white women.'' & 0.5 & 1 & 0 \\
    ``South Africa today is the sad reality of the result of failed colonialism. White South Africans would rather die than share power.'' & 0.53 & 0.8 & 0 \\
    ``Good for the IRR. Maybe those fascist middle class UCT black students - especially that idiotic one with the 'Decolonised Science' rubbish - might actually learn something from this. But I won't be holding my breath - those children wouldn't understand democracy if it jumped up and threw a burning tyre around their necks.'' & 0.85 & 0.4 & 1 \\
    \bottomrule
  \end{tabular}
 \end{table*}
 
While there might be variability within every rater pool along such lines as political affiliation, gender, educational level, etc, the statistically significant difference 
shown in the previous section indicates that there are reasons other than chance for certain differences. While a full qualitative analysis of 25000+ comments is beyond the scope of this paper, we performed content analysis \cite{weber1990basic} of a sample of these comments to understand why some comments are rated differently between control pools and specialized rater pools.

\begin{table*}[ht]
  \caption{Sample comments with themes closely tied to identity where the LGBTQ and control rater pools had high disagreement (high differences in mean scores). We present here the fraction of raters who rated the comment as ``toxic'' or ``very toxic'', -1 or -2 on the Likert scale respectively. This fraction of raters is the value we use for model training, which makes them comparable to Perspective's scores for these comments.}
  \label{tab:example_disagree_comments2}
     \begin{tabular}{p{8cm} | p{1.4cm} | p{1.3cm} | p{1.3cm}}
    \toprule
     Comment & Perspective Toxicity score & Fraction of {\black c}ontrol Pool Raters who rated this toxic & Fraction of LGBTQ Pool Raters who rated this toxic  \\
   \midrule
    ``Here's hoping that their cats' Toxoplasma gondii brain-hijacking parasite causing women to commit suicide nearly two times more often than all other women (whether having been previously diagnosed with emotional or mental problems or not, that does not matter to all T. gondii caused suicides), that their very own cats will be the tipping-point to make them all finally practice what they preach. The fewer of them on the planet the sooner we can bring sustainable ecosystems back into balance for all the sane, responsible, wise, and respectable humans.'' & 0.43 & 0 & 1 \\
    ``I Homosexuality, transgender sex-all are deviations from normal human behavior as well.'' & 0.74 & 0 & 1 \\
    ``Because a homosexual orientation is supposed to not be acted upon whether or not one enters the priesthood? Since a priest sacrifices having a wife and family, how does one sacrifice something they don't desire to begin with?'' & 0.44 & 0 & 0.8 \\
    ``Yeah. Remember that Jesus told us `Love your neighbor as yourself. Unless, of course, if your neighbor is gay; him you shall reject as intrinsically disordered.''' & 0.65 & 0.8 & 0 \\
    ``I am NOT protecting or condoning the alleged behavior!
    
    I'm pointing out the `he did this because he is gay' bigotry.'' & 0.77 & 0.8 & 0 \\ 
    ``I feel the same fear for the gay members of my family.'' & 0.5 & 0.8 & 0 \\
   \bottomrule
  \end{tabular}
 \end{table*}

We created a sample dataset by identifying the top 100 comments with {\black high} mean differences ($\ge\!1$) between the control pool and specialized rater pools. One of the authors next categorized these comments at very high-level domain related themes, like political, religious, gender-based etc. Given the source dataset, it is not surprising that many of the comments are political in nature. So, many of the comments that raters disagreed on were political. When multiple themes were discovered in a comment---the comment was assigned to both the domains. 

By highlighting the comments that were rated significantly differently between control and specialized rater pools in the sections below, we want to share some examples of what kind of content is disagreed upon significantly. We also share how likely Perspective API based ML models consider these comments to be toxic to draw comparison against a well established ML model being used in industry and for academic research.
 
\subsubsection{Mean Differences between Specialized Rater Pools and Control Rater Pool}
We present in \autoref{tab:example_disagree_comments} and \autoref{tab:example_disagree_comments2} a few examples of comments where the rater's African American or LGBTQ identity appears closely tied to the themes presented in the comment. We chose these comments because of their high differences in means between the control and African American/LGBTQ rater groups (they are among the top 100 with the highest differences in means between rater groups). We also consider here the fraction of the rater scores who marked the comment as toxic. The fraction of raters who mark the comment as toxic is the score we use as input for model training, as discussed in the next section. As a point of comparison, we also include Perspective API's scores for these comments.



This matters for two particular cases{\black :} 

\begin{enumerate}
    \item FALSE NEGATIVE CASE: First, where the specialized rater pool rated the comment as more toxic than the control rater pool. If we were only considering ratings from the control rater pool for model building, this could result in models having false negatives--where the comment is actually more toxic than what the model score indicates.
    \item FALSE POSITIVE CASE: Second, where the specialized rater pool rated the comment as less toxic than the control rater pool. If we were only considering ratings from the control rater pool for model building, this could result in false positives--where the comment is actually less toxic than what the model score indicates. 
\end{enumerate}

In both cases, we see that it would be problematic to consider only ratings from the control group, where voices from the minority communities are not heard.

For most comments, the Perspective scores are closer in distance to the control rater pool scores, indicating that the current models of Perspective API could benefit from the inclusion of {\black s}pecialized {\black r}ater {\black p}ools for annotations.

\subsection{Models}
\label{sec:models_eval}

We also wanted to consider how models would perform when trained on the data from the study. We want to understand this because while it's interesting to see differences in the underlying data, the biases observed in prior work were in models, which get trained from the data. Therefore, it is important to test not only the data, but the models themselves. 

To test this, we trained three BERT \cite{devlin2018bert} models (pretrained on data from online conversations) on the data from each of the rater pools.
In \autoref{tab:toxicity_AUCs}, we show the area under the receiver operating characteristic curve (AUC) on three different test sets, a synthetic test set \cite{dixon2018measuring}, the HateCheck dataset from \cite{rottger2020hatecheck}, and the Civil Comments test data \cite{borkan2019nuancedmetrics}. We compare these models to Perspective API's current models \cite{perspective-model-cards}.

\begin{table*}
  \caption{AUC scores for toxicity classifiers trained on data annotated by the three different rater groups (control, LGBTQ, and African American), compared with the Perspective API. The Perspective model is trained on much more data and outperforms our models in most cases; however, the LGBTQ data model is sometimes competitive with it even with much less data. This suggests that training models on data from specialized rater pools can yield surprising performance benefits on test sets considering identities.}
  \label{tab:toxicity_AUCs}
  \begin{tabular}{p{3cm} | p{2cm} | p{2cm} | p{2cm} | p{2cm}}
    \toprule
     Test set & Control model toxicity AUC & LGBTQ data model toxicity AUC & AA data model toxicity AUC & Perspective toxicity AUC
  \\
    \midrule
    Synthetic test set \cite{dixon2018measuring} & 0.973 & 0.987 & 0.972 & 0.994 \\
    HateCheck \cite{rottger2020hatecheck} & 0.746 & 0.792 & 0.709 & 0.664 \\
    Civil Comments \cite{borkan2019nuancedmetrics} & 0.903 & 0.869 & 0.914 & 0.974 \\
    \bottomrule
  \end{tabular}
\end{table*}

Unsurprisingly, the Perspective models usually outperform any of the models trained on data from the experiment, as Perspective has far more training data. However, the LGBTQ data model from the study in this paper is sometimes competitive with this model, indicating that even with less training data, models trained on data from specialized rater pools can have performance benefits on test sets that consider identities. 

We also want to consider whether the models trained on the African American and LGBTQ specialized rater pool data perform differently on identity-related comments than the model trained on the control data. To test this, for the Civil Comments test data, we  analyzed those comments that are tagged with the ``homosexual\_gay\_or\_lesbian'' identity label for which the model trained on the LGBTQ data and the model trained on the control data most disagree. 
We also  analyzed those comments that are tagged with the ``black'' identity label for which the model trained on the African American data and the model trained on the control data most disagree.
These represent examples where model score differences are high, and identities are also mentioned. See some examples in \autoref{tab:model_diff_lgbtq} and \autoref{tab:model_diff_aa}.

\begin{table*}[ht]
  \caption{Sample comments where scores from models trained on data annotated by the control and LGBTQ rater groups differ most.}
  \label{tab:model_diff_lgbtq}
  \begin{tabular}{p{8cm} | p{2cm} | p{2cm}}
    \toprule
     Comment & {\black c}ontrol model score & LGBTQ model score
  \\
    \midrule
    ``As you know, I support the LGBT cause and object to the notion that gay marriage is a sin.'' & 0.520 & 0.115 \\
    ``You and pretty much every person that is in favor of the baker discrimination do not get it.
Political thought is not protected. In Colorado sexual orientation is protected. 

Bottom line?
You do not have to make a cake for a democrat.
You have to make a cake for a gay person.'' & 0.477 & 0.129 \\
    ``How is heavens name does homosexuals doing what for them is perfectly natural `show the failure to love God?''' & 0.385 & 0.733 \\
    ``How about the LGBTQH and P's that should cover all the bases and only the animals can complain that they are left out. I am sure there is some group that feels slighted so lets abolish sports all together.'' & 0.337  & 0.656 \\
    \bottomrule
  \end{tabular}
\end{table*}

\begin{table*}
  \caption{Sample comments where scores from models trained on data annotated by the control and African American rater groups differ most.}
  \label{tab:model_diff_aa}
  \begin{tabular}{p{8cm} | p{2cm} | p{2cm}}
    \toprule
     Comment & {\black c}ontrol model score & African American model score
  \\
    \midrule
    ``Since all black people are the same, then Michelle agrees with all rap lyrics.  Of course!'' & 0.561 & 0.247 \\
    ``Democrats renounced racism. I wish I could say that of Trump's Republicans. More white people get government assistance than black people. Some people develop dependency, but most use it as the safety net it was meant to be. It helps people, and the economy, survive hard times.'' & 0.504 & 0.202 \\
    ``In 2013, whites committed  4,396 homocides, While Blacks committed  5,375 homocides, even though Blacks were  only 13.2\% of the population, and Whites being 64\% of the population.
    - FBI crime statistics 2013.'' & 0.178 & 0.393 \\
    ``In the 500+ days between the Tray von shooting and the Zimmerman trial, more than 11,000 blacks were killed by other blacks, an inconvenient truth for sure?'' & 0.193 & 0.377 \\
    \bottomrule
  \end{tabular}
\end{table*}

We also considered the bias metrics  on a synthetic based test set \cite{dixon2018measuring}. For more details please refer to the Appendix, where these are shown in \autoref{fig:bias_control} for the control data model, in \autoref{fig:bias_lgbtq} for the LGBTQ data model, and in \autoref{fig:bias_aa} for the African American data model.

While deeper qualitative analysis is beyond the goals of this current work, we noticed that when the model results deviated significantly from each other for some comments, certain patterns emerged. The BPSN metric--background-positive, subgroup-negative--was first defined in \cite{borkan2019nuancedmetrics} and represents the ROC AUC calculated on a set of comments containing toxic comments that don't contain the identity term in question (background positive), and non-toxic comments containing the identity term (subgroup negative). As seen in \autoref{fig:bias_control}, the control model performs less well for the identity terms ``gay'', ``queer'', ``homosexual'', and ``muslim'' based on BPSN scores. As seen in \autoref{fig:bias_aa}, the African American annotated model improves the BPSN bias performance for the term ``muslim'' but performance drops for the terms ``gay'' and ``queer''. As seen in \autoref{fig:bias_lgbtq}, the LGBTQ annotated model improves the BPSN bias performance for ``queer'', ``homosexual'', and ``muslim'', and slightly for the term ``gay''. So, in summary, the specialized rater pool data trained models show improvements in BPSN bias scores for several terms over the control data trained models.

\section{Discussion}

A large amount of machine learning research involves the use of crowdworkers, but we as researchers are only now starting to think about how the raters self-identify, and how that might affect the way they annotate toxicity. We discussed earlier the example of African American English; raters who do not identify as African American and are not trained linguists may not have a nuanced understanding of African American English, and that can affect how they rate it for toxicity \cite{sap-etal-2019-risk}.

\subsection{Specialized Rater Pools for Inclusive ML Models}
In this work, we utilize ``{\black s}pecialized {\black r}ater {\black p}ools'': pools of raters crafted based on a particular dimension. In our case, we focus on one part of their identity, either {\black ethnicity} (African American) or sexual orientation/gender-identity  (LGBTQ). With specialized rater pools, we challenge the status quo for who gets to decide what is toxic. This adds an additional question: who is it that gets to decide which models for toxicity are state of the art? As it stands now, published datasets such as Civil Comments \cite{borkan2019nuancedmetrics} and HateCheck \cite{rottger2020hatecheck} are the gold standard, and so the annotators for the datasets are in effect the deciders. Yet specialized rater pools were not used for these annotations, so it is a majority vote from an average sampling of the population that determines which models are more effective at detecting toxicity. But what if members of the affected communities were to be able to make these decisions? After all, they are the ones who are impacted the most by toxicity directed towards their identities. 

Also, we would like to imagine a world in which the gold standard for toxicity datasets was data annotated by specialized rater pools---a concept recently championed by \citet{basile2021toward}. Then we would not only be able to measure which models perform best, but also according \emph{to whom}. This would force researchers and the industry to reckon with the fact that not all identity groups are served equally by existing toxicity classifiers, and would encourage development of models that work well for more communities, as determined by specialized rater pool annotations. 

\subsection{When to Use Specialized Rater Pools?}
In our Results section, we found that raters from specialized rater pools rated comments statistically significantly differently than those from control pools in multiple measures.  In particular, while LGBTQ specialized rater pools rated significantly differently than the control pool in their toxicity score, the African American and control pool showed no significant difference in their toxicity scores. This highlights that there is something about toxicity score and LGBTQ rater pools that is worth further investigation. Is it that the {\black t}oxicity {\black s}core itself is not the best metric to test for identity related differences? Perhaps we should be paying closer attention to other metrics like identity attacks where there w{\black ere} statistically significant difference{\black s} across all the pools. For example, we found that there were higher percentages of identity mentions in comments with high disagreement between the control and specialized rater pools than in comments with low disagreement. More specifically, this means that if specialized and control rater pools disagree on a comment, the comment is more likely to contain an identity reference than if the two rater pools had agreed on the comment. Therefore, we recommend that for comments containing identity references, specialized rater pools should be used to confirm whether or not there is a difference in opinion between groups, which if not addressed could lead to model bias due to the control group's lack of understanding of in-group language around an identity group, or lack of experience being on the receiving end of toxicity towards an identity group.

Often, datasets focused on toxicity and bias contain many identity references \cite{waseem2016racist,borkan2019nuancedmetrics,rottger2020hatecheck}. Therefore, we hypothesize that the use of specialized rater pools to annotate these datasets would produce differing annotations than what is currently published. Given that these datasets are used to determine what is the state of the art for toxicity modeling, we hypothesize that the ``best'' models (as determined today) may not continue to be seen as the best if data was relabeled with specialized rater pools. Further, our work provides a new recommendation for other researchers. Researchers should pay attention to determine if their datasets have identity mentions or attacks. If they do, they should consider using specialized rater pools for data {\black annotation}.

\subsection{Specialized Rater Pools beyond {\black Ethnicity} and Sexual Orientation}
In Section~\ref{sec:methods}, we presented a model of how such pools can be created, and what considerations are necessary when crafting such pools and doing research related to the identity of crowdworkers. We have shown that identity can make a significant impact on the way annotators annotate toxicity, identity attacks, insults, threats, and profanity in text. This opens up the design and empirical space for ML engineers and researchers to use {\black our work} as an example and consider all the different kinds of specialized rater pools we should be imagining, considering, designing, and testing with. {\black It is equally important to perform this work as participatory research in collaboration with all the different rater pools, which we strived for. }

Our goal in this research is to ask researchers and industry professionals to think more critically about who the annotators are for their data and how their identities are impacting the annotations, since this in turn impacts the models they build and models that other researchers build that depend on their data. Using specialized rater pools may enable researchers to build models that work better for more identity groups. While this does not resolve all bias issues (because, for example, members of marginalized communities may still show bias towards their own communities), it still shows a good faith effort towards reducing model bias by directly consulting the experts--the communities themselves.

{\black 
\subsection{Specialized Rater Pools Dataset and Future Directions}
One of the contributions of this work is the creation and publication of the specialized rater pool dataset, as discussed in Section~\ref{sec:dataset}. This consists of 382,500 annotations of 25,500 comments in the Civil Comments dataset by three carefully curated pools of raters that self-identify as LGBTQ, African American, and neither LGBTQ nor African American. We aim to encourage the research community to dig more deeply into Google Specialized Rater Pools Dataset at 
\url{https://www.kaggle.com/datasets/google/jigsaw-specialized-rater-pools-dataset}  {\footnote{Nitesh Goyal, Ian Kivlichan, Rachel Rosen, \& Lucy Vasserman. (2022). [Data set]. Kaggle. https://doi.org/10.34740/KAGGLE/DSV/3533200}}.

and perform subsequent analysis into when and how identity-based specialized rater pools might be integrated into ML development. As a starting point, this could include replicating our results. We additionally hope that the community will use this dataset as an opportunity to expand upon this work by imagining and creating their own specialized rater pools across other identities.

Further, we performed preliminary content analysis in the paper. Our initial findings suggest that the identity attack, threat, and profanity categories are the differentiating reasons for different ratings between the control pool and specialized rater pools. Qualitative researchers can choose to perform deeper linguistic and content analysis to identify the specific markers in language that lead to these differences. Practitioners can use this dataset and approach to compare how their existing datasets will perform when rated by specialized rater pools. 

Finally, we expect this dataset to be useful for the machine learning community. In lieu of identity group information, several recent works have constructed simulated groups using methods like graph clustering on the basis of annotators' responses \cite{akhtar2020modeling, basile2020s, wich2020investigating}. The dataset we are releasing here makes this information accessible, enabling further evaluation and comparison of these methods with the underlying identity groups. Additionally, ML practitioners could further analyze the predictions generated by models trained on different identity pools' annotations, expanding upon our work in Section~\ref{sec:models_eval}. For example, the data in \citet{sap-etal-2019-risk} could be further evaluated using models trained on data from these three specialized rater pools so as to further understand different sources of potential bias. Overall, we hope that this specialized rater pool dataset will create new opportunities and research directions for academics and practitioners alike.
}

\section{Limitations}
The specialized rater pools in this study were limited to African Americans and LGBTQ Americans, but future research should be expanded to include other groups as well to explore if these findings would hold true for other identity-related rater pools. Similarly, even though the comments should be rated by the group that is targeted, we need more research to understand what happens when different pools disagree. This becomes increasingly important when working with intersectional identities. This work focused on binary identities, and future work should go beyond binary identities.

Future work could also include more community-specific data sources to look at community-specific discussion, instead of discussions about communities, to be more reflective of {\black the} relationship between conversations and the identity{\black -}related linguistic markers. Future work should also consider deeper qualitative research methods to fully understand how and why {\black the} data differ{\black s}.

\section{Conclusion}

The intent of the research was to determine if specialized rater pools, made up of individuals who identify as African American or LGBTQ American, rate comments differently than control rater pools made up of individuals who don't identify as these identities. We show this to be true {\black in several cases}.
We also look at the performance on models trained on data from the study and show that even models trained on smaller {\black datasets labeled by} specialized rater pools can perform better than models {\black trained} on larger datasets labeled by randomly assigned annotators. 

\begin{acks}
We thank Olivia Redfield and Raquel Saxe for contributions to the early stages of this work. We also would like to thank Alyssa Whitlock Lees, Jeffrey Sorensen and other anonymous reviewers for suggestions on improving this work.
\end{acks}

\bibliographystyle{ACM-Reference-Format}
\bibliography{sample-base}
\newpage
\appendix


\section{Rater Template}

\begin{figure}[H]
  \includegraphics[width=\linewidth]{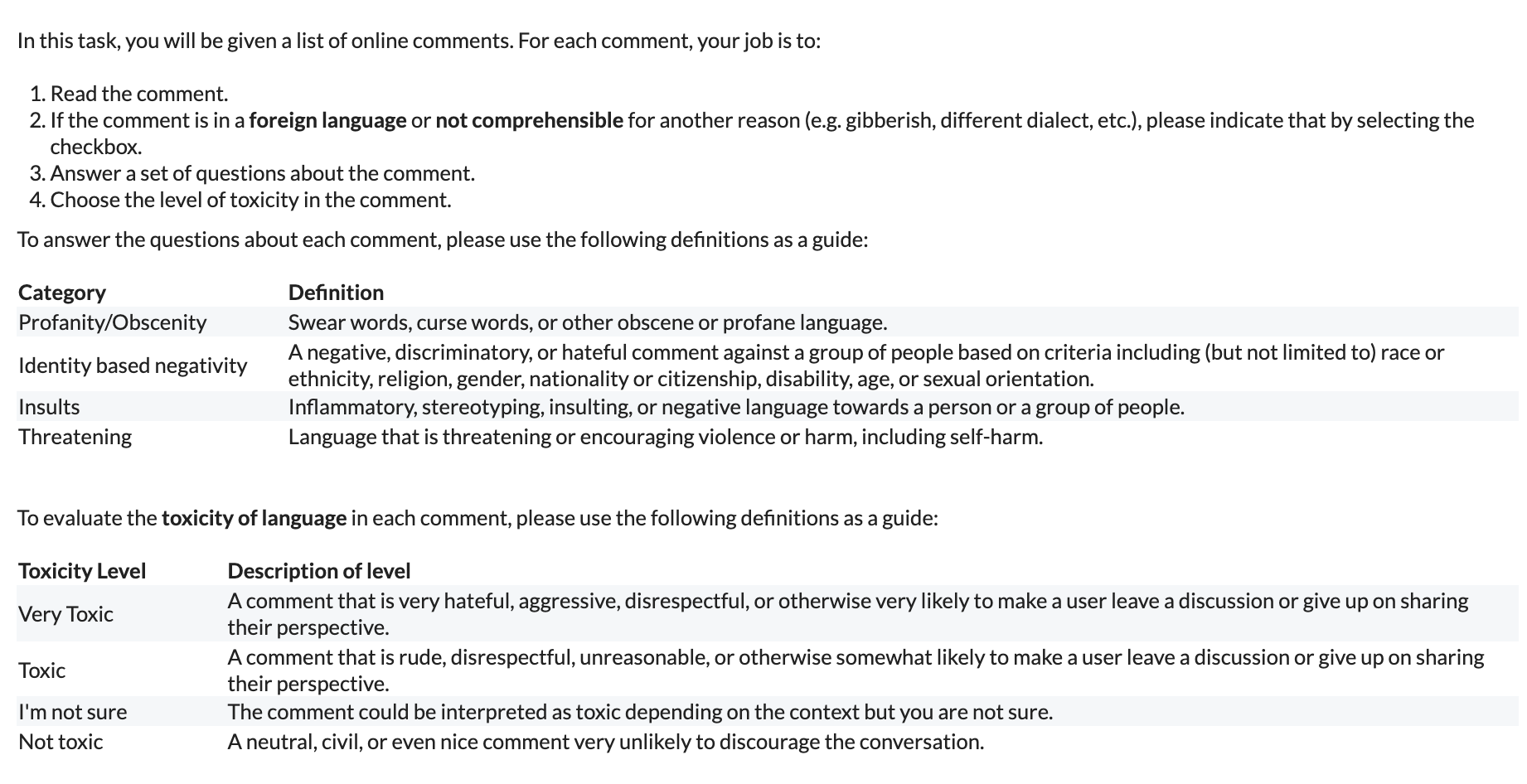}
  \includegraphics[width=\linewidth]{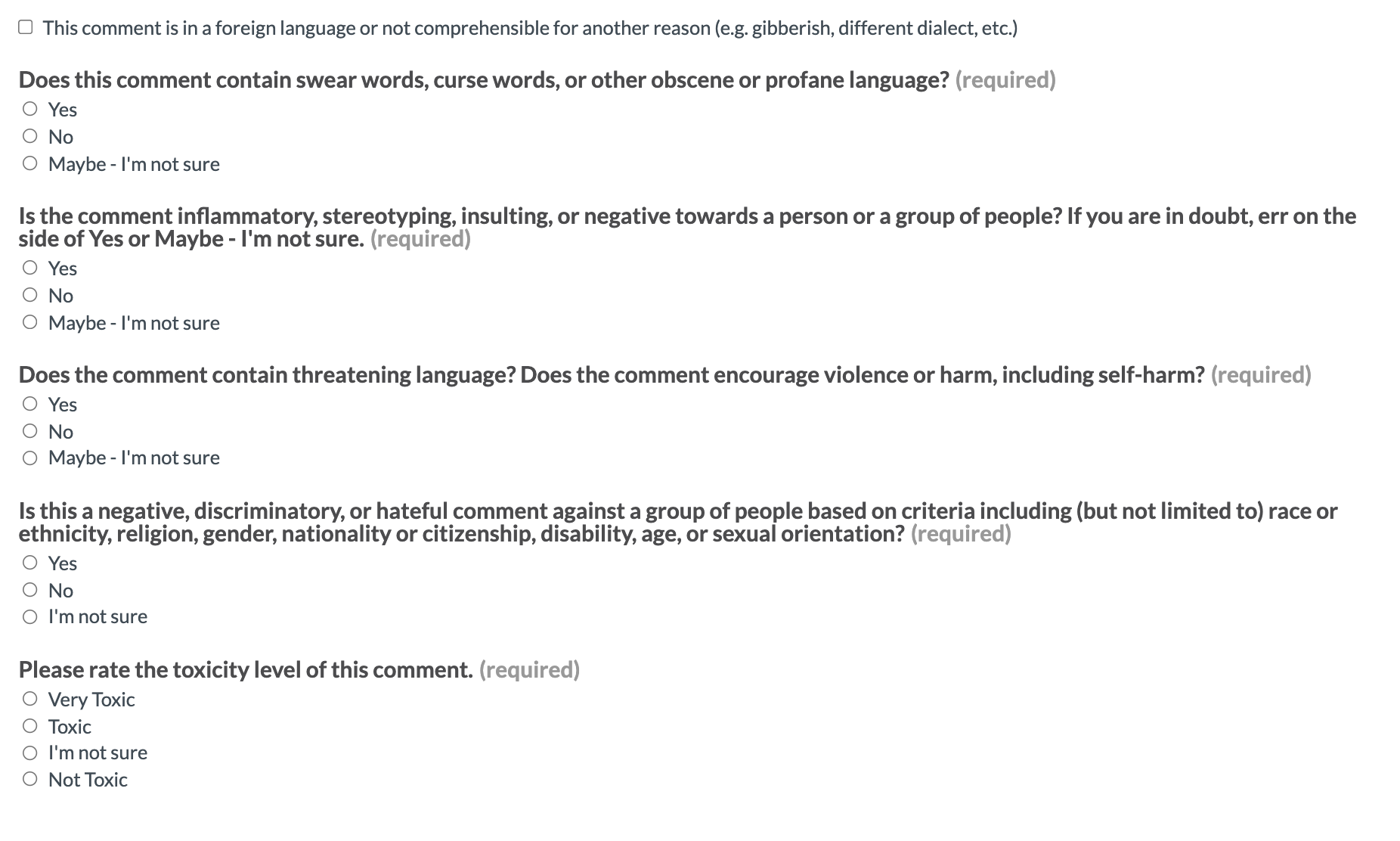}
  \caption{Rating template used for this work. \label{fig:template}}
  \Description{A sample rating template as used for the study. The template instructs annotators how to categorize a list of online comments and choose a toxicity level for both, including defining the different categories (profanity, identity based attacks, insults, and threatening) and describing the toxicity levels, from very toxic to not toxic.}
\end{figure}

\begin{figure}[H]
  \includegraphics[width=\linewidth]{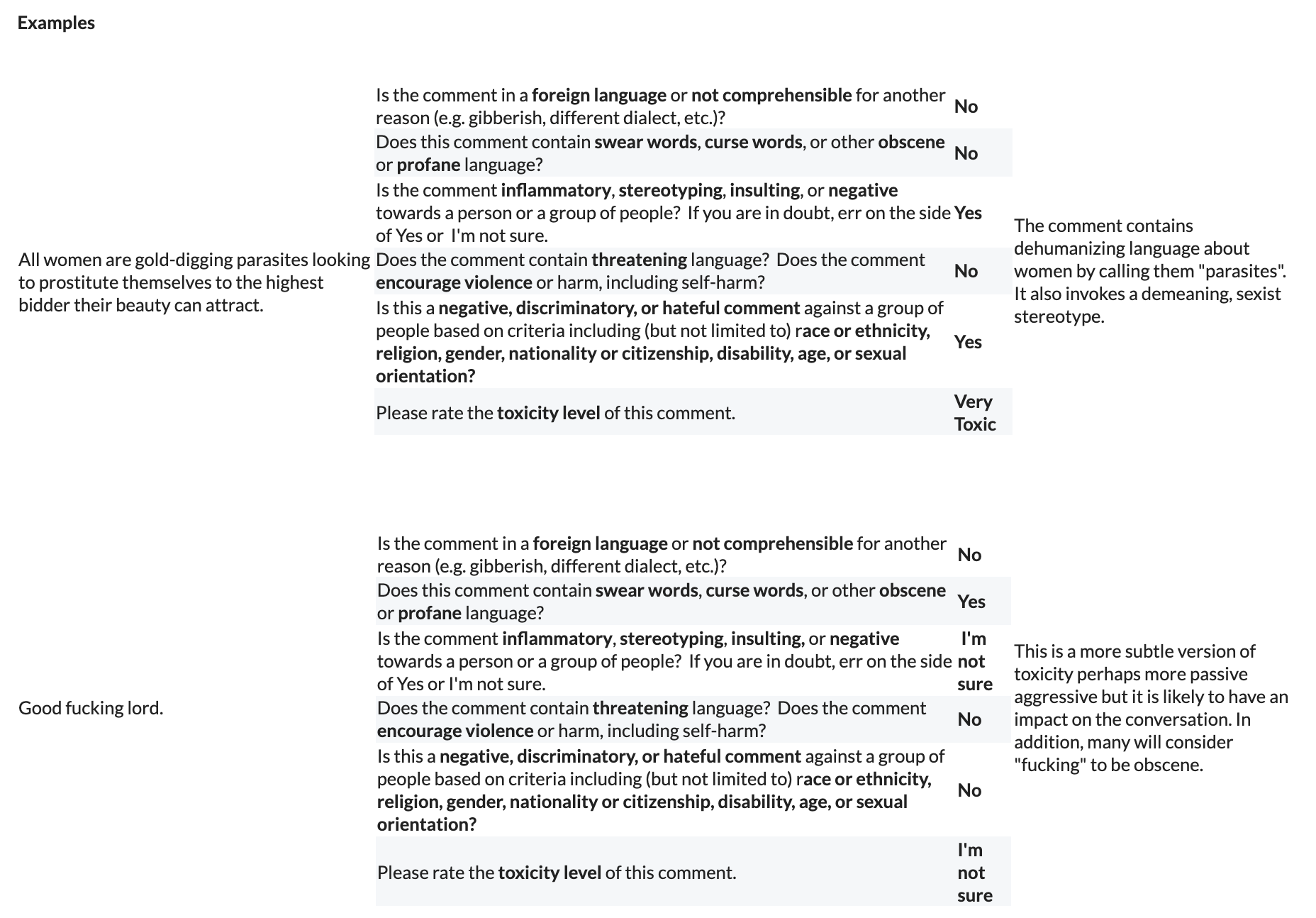}
  \includegraphics[width=\linewidth]{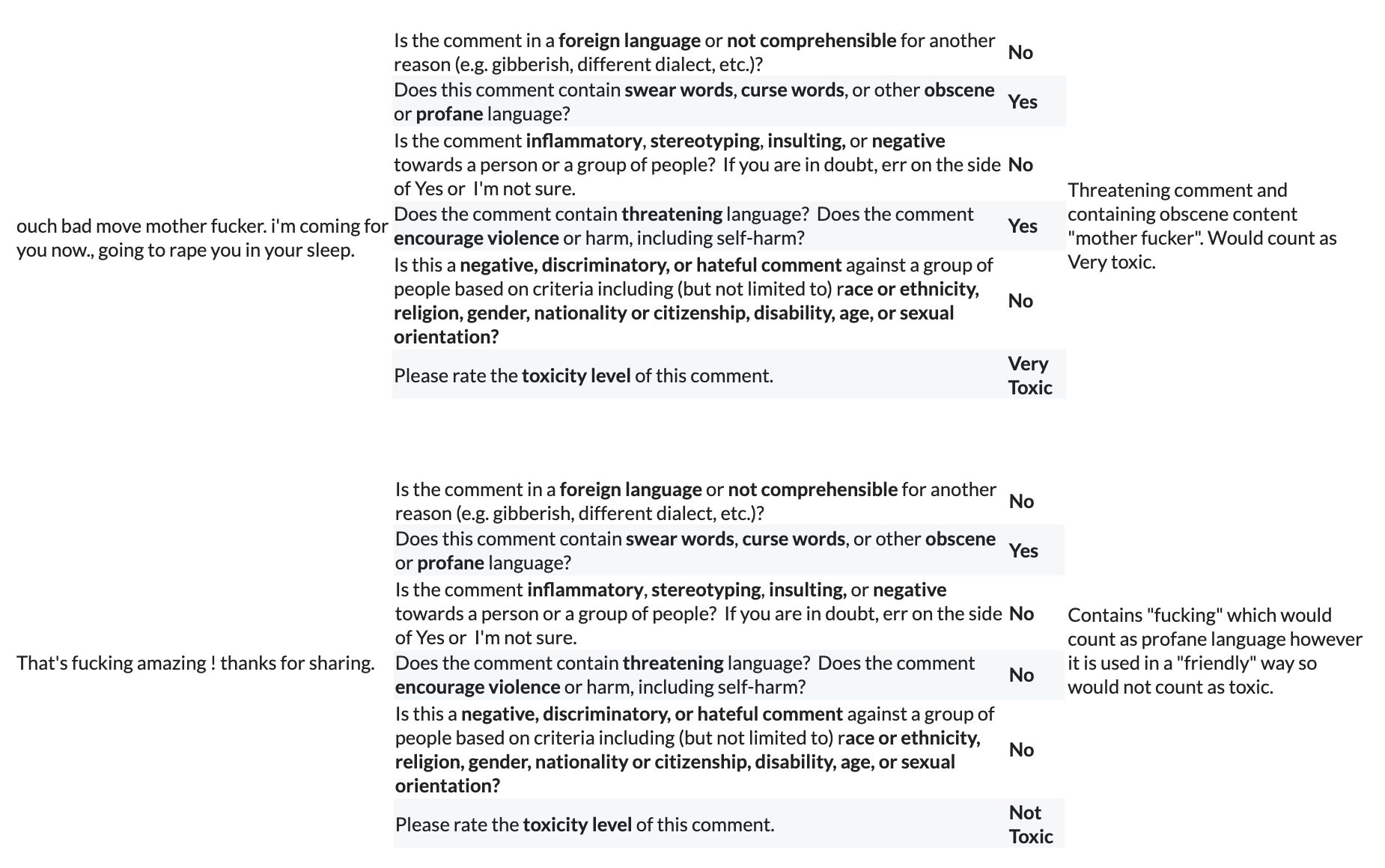}

  \includegraphics[width=\linewidth]{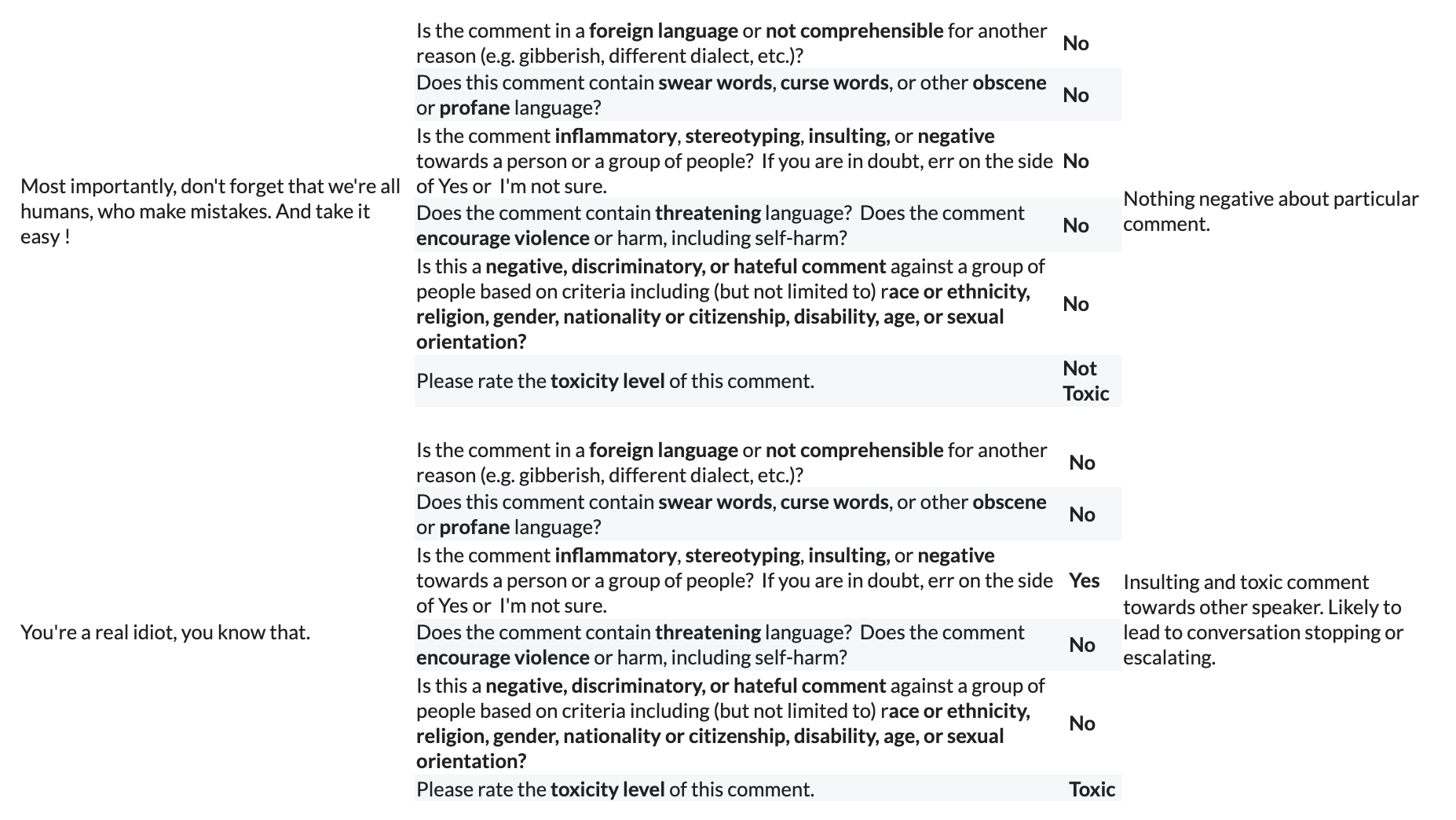}
  \includegraphics[width=\linewidth]{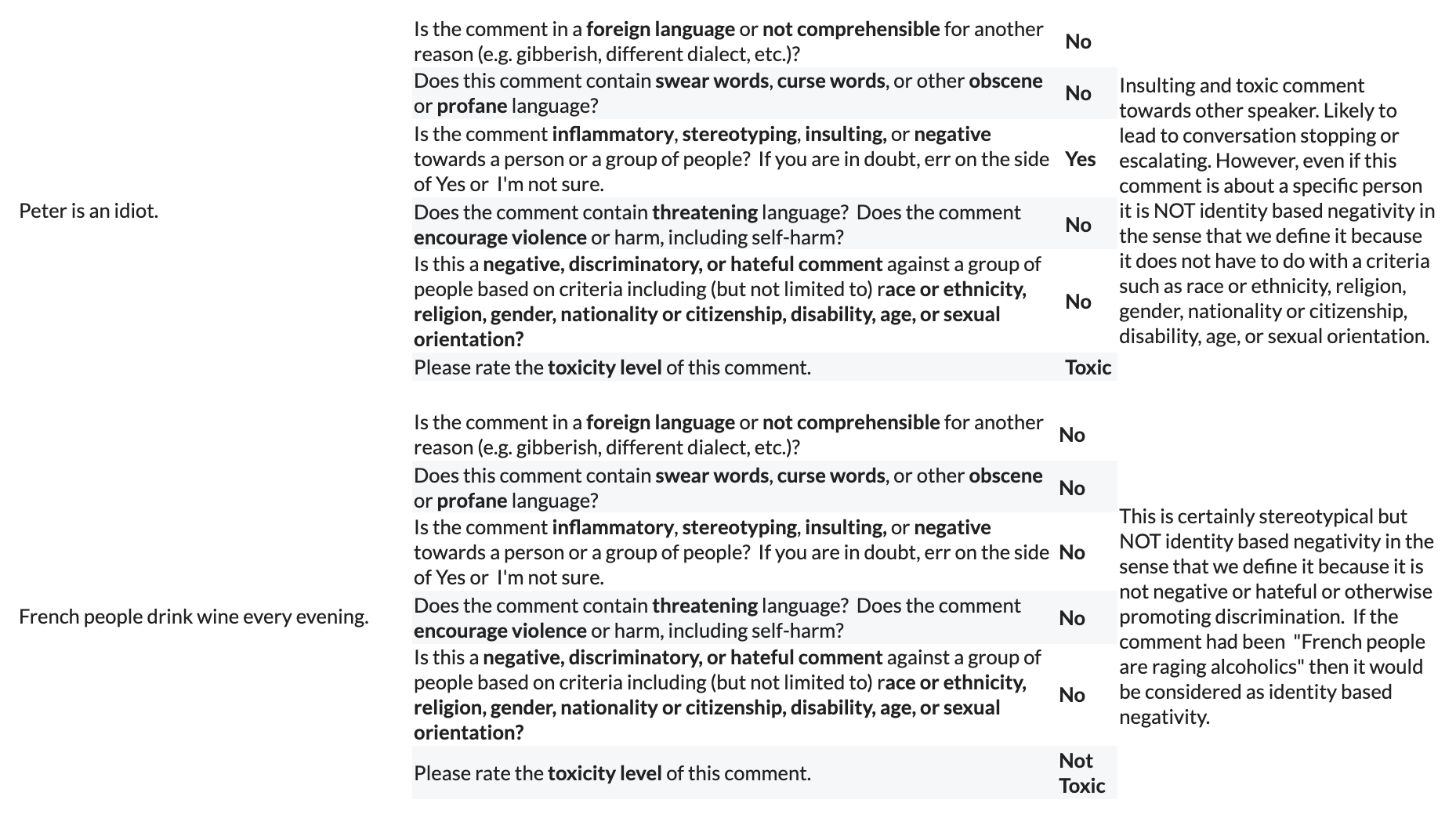}
  \caption{Examples given for the rating instructions for this work. \label{fig:template_examples}}
  \Description{Examples given to raters for the instructions in the study.}
\end{figure}

In this appendix, we include the rating template (\autoref{fig:template}) and examples for the rating instructions (\autoref{fig:template_examples}) used for this work.

\section{Model Bias Metrics}

\begin{figure}[H]
  \includegraphics[height=0.71\textheight]{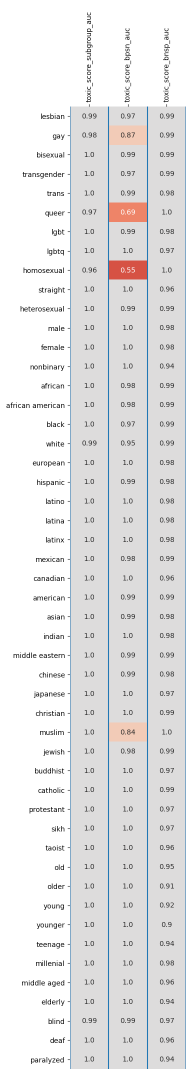}
  \caption{Bias metrics for the control data model. The BPSN metric--background-positive, subgroup-negative--represents the ROC AUC calculated on a set of comments containing toxic comments that don't contain the identity term in question (background positive), and non-toxic comments containing the identity term (subgroup negative). As seen here, the control model has lower BPSN scores for the identity terms ``gay'', ``queer'', ``homosexual'', and ``muslim''. \label{fig:bias_control}}
\end{figure}

\begin{figure}[H]
  \includegraphics[height=0.71\textheight]{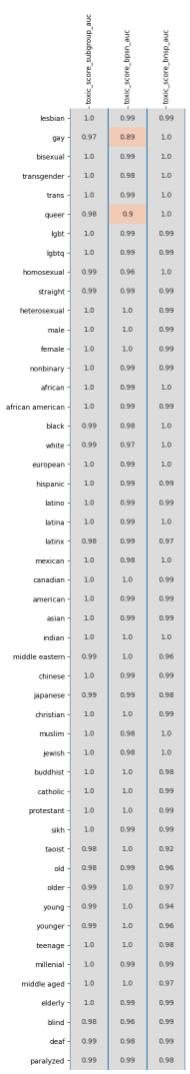}
  \caption{Bias metrics for the LGBTQ data model. The BPSN metric--background-positive, subgroup-negative--represents the ROC AUC calculated on a set of comments containing toxic comments that don't contain the identity term in question (background positive), and non-toxic comments containing the identity term (subgroup negative). As seen here, the LGBTQ annotated model improves the BPSN bias performance for ``queer'', ``homosexual'', and ``muslim'', and slightly for the term ``gay'' compared to the control model.  \label{fig:bias_lgbtq}}
\end{figure}

\begin{figure}[H]
  \centering
  \includegraphics[height=0.71\textheight]{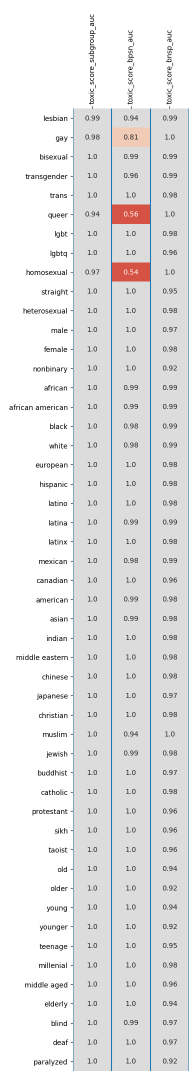}
  \caption{Bias metrics for the African American data model. The BPSN metric--background-positive, subgroup-negative--represents the ROC AUC calculated on a set of comments containing toxic comments that don't contain the identity term in question (background positive), and non-toxic comments containing the identity term (subgroup negative).  As seen here, the African American annotated model improves the BPSN bias performance for the term ``muslim'' but sees a drop in performance for the terms ``gay'' and ``queer'' compared to the control model. } \label{fig:bias_aa}
\end{figure}

We list bias metrics for the three models we trained on data from the three different specialized rater pools. \autoref{fig:bias_control} lists metrics for the control pool model, \autoref{fig:bias_lgbtq} for the LGBTQ pool model, and \autoref{fig:bias_aa} for the African-American pool model.

\end{document}